# Securing Remote Procedure Calls over HTTPS

Ajinkya Kale, Ashish Gilda, Sudeep Pradhan
*Department of Computer Engineering & Information Technology,*
*College of Engineering, Pune-05*
kaleajinkya@gmail.com , ashishpgilda@gmail.com , sudeepster@gmail.com

**Abstract**
*Remote Procedure Calls (RPC) are widely used over the Internet as they provide a simple and elegant way of interaction between the client and the server. This paper proposes a solution for securing the remote procedure calls (RPC) by tunneling it through HTTPS (Hypertext Transfer Protocol over Secure Socket Layer). RPC over HTTP actually uses the Secure Socket Layer (SSL) protocol as a transport for the traffic. SSL mandates that the server authenticates itself to the client using a digital certificate (and associated private key). SSL is normally configured to encrypt traffic before transmitting it between the server and client and vice versa.*

## 1. Introduction

"RPC over HTTPS" tries to provide a solution to secure the Remote Procedure calls made by a client outside a network to access the Server services. These services can now be accessed via internet securely using the https protocol.
Remote Procedure Call data or traffic will be encapsulated in http protocol so that the data can pass through internet services. Using HTTP with SSL i.e. HTTPS makes the authentication and authorization easy and adds security to the traffic.

## 2. Issues using current RPC implementation

While RPC provides an elegant mechanism for client-server interaction, it suffers from the following problems:

**2.1.** RPC usually uses TCP or UDP directly and different RPC servers use different port numbers. Site administrators usually firewall access to these non-standard ports and this prevents use of RPC on the internet to talk to servers. Site administrators like to open port access to only a few well tested services like http and ftp, due to the scare of possible vulnerabilities in new services.

**2.2.** RPC traffic is not secure, that is, it is not confidential and tamper-proof. RPC messages can be snooped and modified without getting noticed. It suffers from common security problems such as authentication and authorization.

## 3. Proposed Solution

Since HTTP is a widely used and trusted protocol with almost all sites allowing it to pass through their firewalls, encapsulating RPC in http would solve **problem 1 stated in Section 2.2** Using HTTPS would solve **problem 2 stated in Section 2.2** The resulting communication will have all the properties of the secured protocol and will in turn secure the Remote Procedure Calls.
By doing this as a generic RPC encapsulating facility, any client/server application that uses RPC will immediately benefit from it. Some of the applications are – NFS, Microsoft Exchange/Outlook and other applications listed above. Thus all these applications will become immediately accessible over the internet.

## 4. Related Work [2]
. As of now the only application that wraps the Remote Procedure Calls in Text protocols such as HTTP/HTTPS is the Microsoft Outlook(2003).
The RPC protocol allows Outlook 2003 MAPI clients to connect to Exchange 2003 Servers. Because of very restrictive firewall rules which typically allow port access for only HTTP; this prevents remote users from accessing Exchange Servers directly. Blocking secure RPC connections prevents your remote users from benefiting from the full Outlook MAPI client.

Microsoft realized the magnitude of this problem. Their solution is the RPC over HTTP protocol. This protocol allows remote Outlook 2003 clients to connect to Exchange 2003 Servers using HTTP or HTTPS. The RPC protocol commands and data are "wrapped" (as known as *encapsulated*) in an HTTP header. The firewall in front of the Outlook 2003 MAPI client only sees the HTTP header and passes the outbound connection through. The RPC over HTTP protocols allows your remote users to get around what might be considered an overly zealous approach to outbound access control.

However, Microsoft's solution is not generalized for other RPC applications. There are no open source implementations available for such wrapping and security of the RPCs. Hence our project will provide a generalized solution for the security issues for applications using RPCs.

## 5. Using HTTP as RPC Transport

RPC-over-HTTP enables client programs to use the Internet to execute procedures provided by server programs on distant networks. RPC over HTTP tunnels its calls through an established HTTP port. Thus, its calls can cross network firewalls on both the client and server networks.

RPC over HTTP routes its calls to the RPC proxy located on the RPC server's network. The RPC Proxy establishes and maintains a connection to the RPC server. It serves as a proxy, dispatching remote procedure calls to the RPC server and sending the server's replies back across the Internet to the client application.

When the client program issues a remote procedure call using HTTP as the transport, the RPC run-time library on the client contacts the RPC proxy. Depending on whether the RPC client was asked to use HTTP or HTTPS (HTTP with SSL) port 80 or port 443 is used, respectively. The RPC proxy contacts the RPC server program and establishes a TCP/IP connection. The client and the RPC proxy maintain their HTTP or HTTPS connection across the Internet. The client's HTTP or HTTPS connection to the RPC proxy can pass through a firewall (subject to appropriate access permissions) if one is present. The server can then execute the remote procedure call and use the connection through the RPC proxy to reply to the client.

If either the client or the server disconnects for any reason, the RPC proxy will detect it and end the RPC session. As long as the session continues, the RPC proxy will maintain its connections to the client and the server. It will forward remote procedure calls from the client to the server, and send replies from the server to the client.

## 6. RPC over HTTPS security

RPC over HTTPS provides three types of security in addition to standard RPC security, which results in RPC over HTTPS traffic being protected once by RPC, and then doubly protected by the tunneling mechanism provided by RPC over HTTPS. The RPC over HTTPS tunneling mechanism adds to normal RPC security in the following manner:
1. Provides SSL encryption and RPC Proxy verification (mutual authentication).
2. Provides restrictions on the RPC Proxy level dictating which machines on the server network are allowed to receive RPC over HTTPS calls.

## 7. Implementation Details

**TCPFilter**

The tcpfilter program sits between the client like a web browser and the server to filter the data transfer between the client and the server.

The requests sent by the client ie client data is served by the filter and passed to the server. The server data written to a particular port can be sent to the client through the tcpfilter.

Tcpfilter is a in between program which acts as a server for the client and a client for the server at the same time.

For a web browser tcpfilter is run using the following format:
tcpfilter <source port> <destination machine> <destination port>

eg: tcpfilter 80 192.168.0.1 100

(In the above example a server program is running on 192.168.0.1 which dumps data on port 100. Client ie web browser gets the data from the port 100 through the tcpfilter)

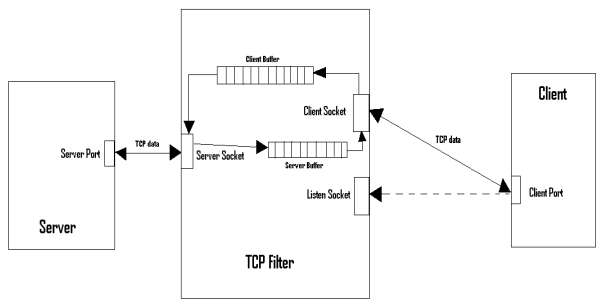

**Figure 1: TCPFilter**

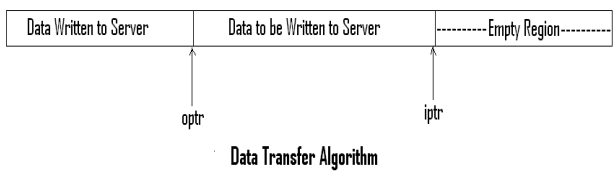

**Figure 2 : Data Transfer Algorithm**

**Portmapper:**
Once the tcp data was transferred through the tcpfilter the need was to get the rpc traffic between client and server through the filter.

The portmapper utility gives us the methods to register the program number and version number pair to the specific port.

This module registers the specific port numbers to the "nfs" and "mountd" programs.

Nfs program number: 100005  Version numbers : 1 , 2 , 3

Mountd program number: 100003  Version numbers : 2 , 3 , 4

This registering ensures that the filter sends the data to the specific code which in turn is sent to the nfs and mount daemons.

**RPCFILTER**
RPCFilter combines the registering program and the TCPFilter into one module which helps to filter the rpc traffic between client and server.

Now RPC call requests like mount are serviced by the rpcfilter. RPCFilter send the rpc call request packets from client to the nfs server machine using the registered ports. The server services the request and sends the rpc response packets back to the rpcfilter which in turn send the response packets to the client where the rpc call was made. This completes one RPC Request-Response cycle.

RPCFilter program is run at the client side. The invoking format is as follows:

rpcfilter <server machine address> <program number> <version numbers>

Every time 2 instances of the filter are invoked at the client, one for nfs (100005) and another for mountd (100003).

rpcfilter   192.168.0.1   100005   1   2   3

rpcfilter   192.168.0.1   100003   2   3   4

Where 192.168.0.1 is the address of the nfs server machine.

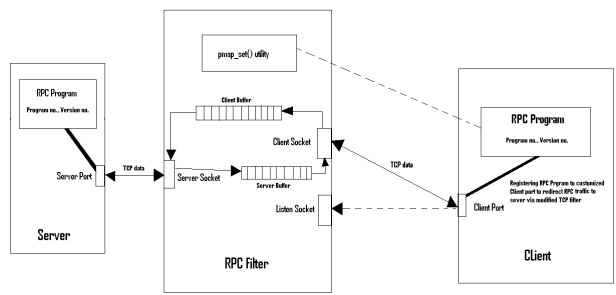

**Figure 3 : RPCFilter**

**Server Module**:

The Common Gateway Interface (CGI) is a standard for interfacing external applications with information servers, such as HTTP or Web servers. A plain HTML document that the Web daemon **retrieves** is **static**, which means it exists in a constant state: a text file that doesn't change. A CGI program, on the other hand, is **executed** in real-time, so that it can output **dynamic** information.

We use the property of CGI to execute at real time for decapsulating RPC(binary) data from HTTP which is a text protocol. The RPC data is POSTed by the client (rpcFilter) to the web server(Apache HTTPD) by setting the content type to application/stream. Apache web server removes the POST headers in the byte stream from client (rpcFilter) and makes the byte stream available to the standard input of the CGI bin program. Hence we get the entire RPC call at the standard input of the CGI bin program which is stored in the input buffer.

The CGI program gets the program number(prog) and version number(vers) by extracting the appropriate unsigned 32-bit integers. The program number and version number are used to get the port number(portnum) which gives the port on which the RPC program is running on local machine. The CGI program uses the pmap_getport() function in <rpc/rpc.h> to get the port number.

The CGI program uses the port number on which the required RPC program is running to create a socket on local machine. The CGI program connects this socket to the port of RPC program. Now the entire input buffer is **written** to the socket. The RPC RFC doesn't distinguish between different machines on the network. So the RPC call to the same program and version with same parameters will have identical RPC call records. This property is exploited by writing the entire RPC call record to the socket connecting the RPC program. This means CGI program is now making a RPC call to the local machine.

The CGI program then receives the response from the RPC program on the same socket. It **reads** the response record from the RPC program into the output buffer. The CGI program now adds HTML header and footers to the output buffer and sends the data to its standard output. Apache web server sends this RPC response encapsulated in HTTP to the client (rpcFilter).

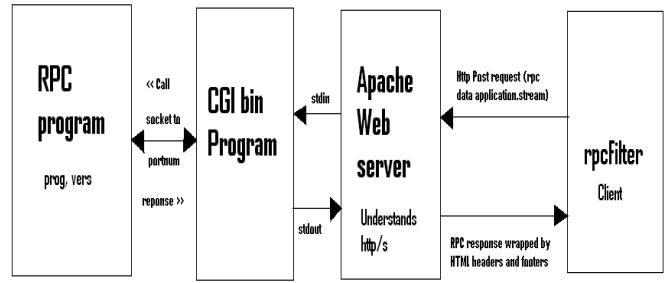

## Conclusion

A generalized solution for securing Remote Procedure Calls over HTTP was implemented. With this solution a client can make RPC calls securely on servers behind Firewalls. The HTTPS protocol introduces security services such as authentication and authorization to RPC call response mechanism. An Open Source solution for accessing Server services over internet (using https) which bypasses the need for a Virtual Private Network was provided.